\documentclass[]{spie}  

\usepackage{enumerate} 
\usepackage{amsmath,amsfonts,amssymb}
\usepackage{graphicx}
\usepackage[colorlinks=true, allcolors=blue]{hyperref}
\usepackage{multirow}
\usepackage{booktabs}
\usepackage{xcolor}
\title{A cascaded deep network for automated tumor detection and segmentation in clinical PET imaging of diffuse large B-cell lymphoma}

\author[1, 2]{Shadab Ahamed}
\author[1]{Natalia Dubljevic}
\author[4]{Ingrid Bloise}
\author[5]{Claire Gowdy}
\author[4]{Patrick Martineau}
\author[4]{Don Wilson}
\author[3, 4]{Carlos F. Uribe}
\author[1, 2, 3]{Arman Rahmim}
\author[2]{Fereshteh Yousefirizi}
\affil[1]{Department of Physics \& Astronomy, University of British Columbia, Vancouver, BC, Canada}
\affil[2]{Department of Integrative Oncology, BC Cancer Research Institute, Vancouver, BC, Canada}
\affil[3]{Department of Radiology, University of British Columbia, Vancouver, BC, Canada}
\affil[4]{BC Cancer, Vancouver, BC, Canada}
\affil[5]{BC Children’s Hospital, Vancouver, BC, Canada}

\authorinfo{Further author information:\\S. A.: E-mail: shadab@phas.ubc.ca}

\begin{document} 
\maketitle
\begin{abstract}
Accurate detection and segmentation of diffuse large B-cell lymphoma (DLBCL) from PET images has important implications for estimation of total metabolic tumor volume, radiomics analysis, surgical intervention and radiotherapy. Manual segmentation of tumors in whole-body PET images is time-consuming, labor-intensive and operator-dependent. In this work, we develop and validate a fast and efficient three-step cascaded deep learning model for automated detection and segmentation of DLBCL tumors from PET images. As compared to a single end-to-end network for segmentation of tumors in whole-body PET images, our three-step model is more effective (improves 3D Dice score from 58.9\% to 78.1\%) since each of its specialized modules, namely the slice classifier, the tumor detector and the tumor segmentor, can be trained independently to a high degree of skill to carry out a specific task, rather than a single network with suboptimal performance on overall segmentation.   
\end{abstract}

\keywords{Diffuse large B-cell lymphoma, $^{18}$F-FDG PET, deep learning, classification, detection, segmentation}

\section{INTRODUCTION}
\label{sec:intro}  

Diffuse large B-cell lymphoma (DLBCL) is the most common form of non-Hodgkin lymphoma among adults \cite{pmid9166827}. Accurate detection and segmentation of DLBCL tumors from $^{18}$F-FDG PET images has important implications for surgical intervention and radiation therapy \cite{pmid33315178, pmid28971313}. Tumor segmentation is required to calculate the total metabolic tumor volume (TMTV); a metric that has been shown to have predictive value for patient outcome in lymphoma \cite{pmid32532925, pmid30625203, pmid31978225, pmid32724136, tmtv_1}. Manual segmentation of tumors in whole-body PET images is time-consuming, labor intensive and operator-dependent, with high intra- and inter-operator variability, hence not performed routinely \cite{pattern_recog}. 

With the advent of more computational power and various breakthroughs in the development of AI algorithms, supervised deep learning-based methods are being increasingly used for localizing the regions of interest (ROIs) in PET images. Unfortunately, these supervised methods require a large amount of accurately annotated data to learn to perform significantly accurate predictions of segmentation contours \cite{pmid32068507, LUNDERVOLD2019102}. Their performance can, in fact, be particularly challenging, especially when the disease is generalized, or the tumors are small and occur in close proximity to one another. Furthermore, automated lesion detection/segmentation based on traditional machine learning techniques have not found widespread use in clinics due to limitations in performance, reproducibility, and explanability \cite{pmid33771905}.

Our overarching aim is to enable routine detection and segmentation of all tumors in PET images of DLBCL patients. In this study, we have implemented a deep learning-based, fully-automated three-step network architecture that takes a 3D PET image as input and outputs the segmentation contours for each of the 2D axial slices containing tumors. This three-step model also improves the Dice score between the ground truth and the predicted segmentation contours on the DLBCL test set by about 19\% as compared the Dice score evaluated using a single end-to-end segmentation network (U-Net) applied on the whole-body PET image.

\begin{figure}

\begin{center}
\includegraphics[scale=0.24]{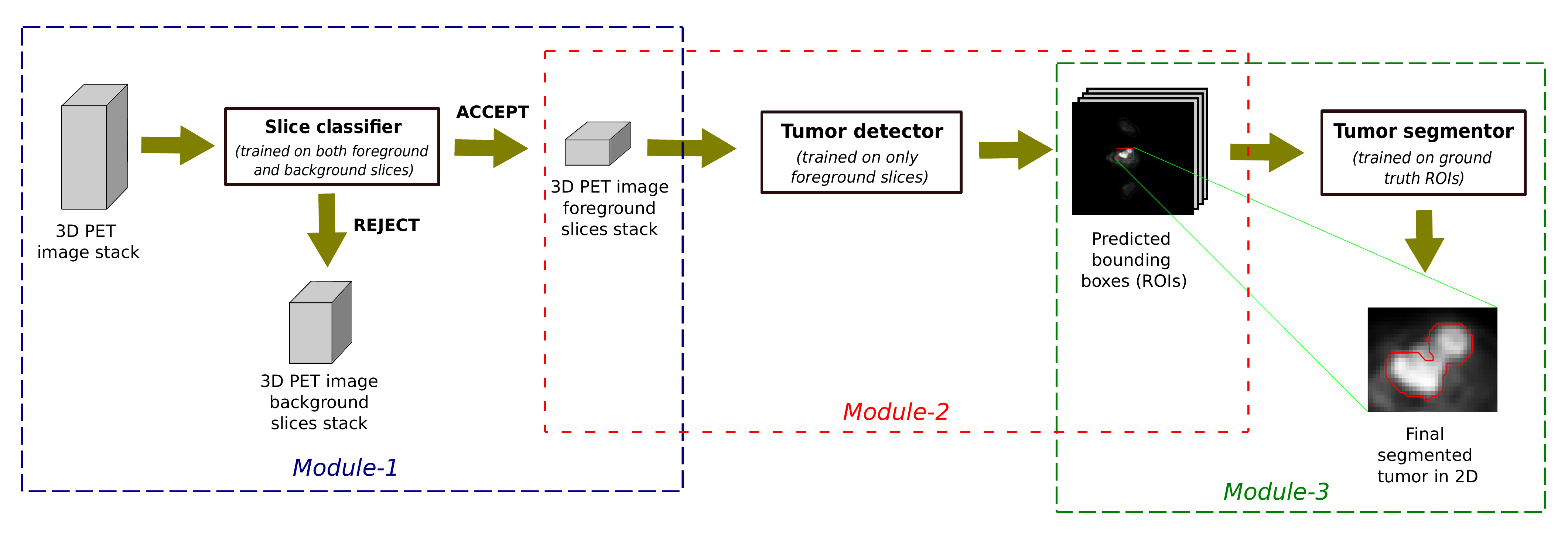}
\end{center}
\caption{The proposed cascaded 3-step deep network for PET tumor segmentation. Module-1 (Slice classifier) is a ResNet152 based binary classification network trained on all axial PET slices. Module-2 (Tumor detector) is a Faster R-CNN based object detection network trained only on the axial foreground PET slices. Module-3 (Tumor segmentor) is a 2D U-Net network trained on ROIs extracted from the ground truths of Module-2.} 
\label{fig:network_arch}
\end{figure} 

\section{METHODS}
\label{sec:methods}
Object detection in medical images refers to identifying location of tumors/organs (objects) via the prediction of bounding boxes around them and classifying different objects. Lesion/tumor segmentation in medical images involves the prediction of the boundaries of lesions/tumors, which is in an essential step towards the calculation of TMTV. There are various approaches to performing fully-automated supervised deep learning-based segmentation of clinical PET images \cite{pmid34537131}. Some approaches employ a single-step method via a single end-to-end segmentation network applied on whole-body images such as U-Net \cite{unet}. Others employ a two-step approach, where the detection step is performed along with the segmentation step within the detected bounding box, either simultaneously \cite{maskrcnn, pmid31217099, pmid29652908,det_seg_1} or one after the other \cite{5459262, pmid33315178, pmid33937842}. In this section, we give a detailed description of our 3-step network architecture and the lymphoma datasets used for training and validation of the model.

\subsection{Structure of the proposed cascaded deep network}
\label{subsec:networks}
Our proposed cascaded segmentation pipeline (Figure \ref{fig:network_arch}) consists of the following three consecutively stacked modules: 

\begin{enumerate}
\item Module-1 (Slice classifier): This module classifies the axial slices of 3D PET images into slices that contain tumors (foreground slices) and slices that do not (background slices). This module is a binary classification network that uses a ResNet152 \cite{7780459} feature extractor (backbone) pre-trained on the ImageNet dataset \cite{5206848}. This backbone is followed by a fully connected layer ending with a sigmoid activation function to facilitate binary classification. The axial slices that contain tumors are given as input to Module-2.

\item Module-2 (Tumor detector): This module localizes tumors on the axial slices obtained from Module-1 by generating rectangular bounding boxes around the suspicious regions. This module uses the object detection network Faster-RCNN \cite{pmid27295650} with ResNet50 feature extractor and an associated feature pyramid network \cite{fpn}, pre-trained on the ImageNet dataset. The coordinates of these bounding boxes are given as input to Module-3.

\item Module-3 (Tumor segmentor): This module segments the tumor inside the rectangular bounding boxes obtained from Module-2. It uses a conventional 2D U-Net architecture \cite{unet} modified for smaller images with a final sigmoid activation layer for the pixel-wise classification into foreground and background pixels.
\end{enumerate}

In this work, we have employed a three-step model instead of directly segmenting the tumors in 3D PET images. The motivation behind this are as follows: (i) In the first step, we train the slice classifier network (Module-1) to learn to reject background slices and accept only the foreground slices, which are subsequently given as input to the tumor detector (Module-2). Rejecting the background slices before the tumor detection step is beneficial as it reduces the dependence of training the detection module on the background slices that contain no tumors or suspicious ROIs. (ii) Furthermore, we hypothesize that segmenting tumors inside the detected bounding boxes improves the segmentation performance by centering the object of interest (a DLBCL tumor) in the receptive field of view and cropping out extraneous information such as normal active organs and other high uptake regions that may degrade the performance.

\subsection{Dataset}
\label{subsec:dataset}
Our lymphoma PET images dataset consisted of 126 cases of primary mediastinal large B-cell lymphoma (PMBCL) (for pre-training) and 50 cases of DLBCL. All images were annotated by nuclear medicine physicians, following initial discussions and development of consensus procedures \cite{consensus}. We used a 60\%:20\%:20\% splitting of these axial slices for training, validation and testing, respectively. Class imbalance, i.e., proportion of slices with tumors (foreground slices) and slices without tumors (background slices) for both these datasets was about 10\%:90\%, across all training, validation and testing sets. Across the three sets, we had a total of 31126 axial slices (2958 foreground and 28168 background) of PMBCL cases and 13185 axial slices (1217 foreground and 11968 background) of DLBCL cases. The axial slices of the PMBCL cases were either 168$\times$168 pixels (pixel dimension: 4.06 mm$\times$4.06 mm) or 192$\times$192 pixels (pixel dimension: 3.64 mm$\times$3.64 mm) in dimension, while the axial slices of the DLBCL cases were 192$\times$192 pixels (pixel dimension: 3.64 mm$\times$3.64 mm). All the axial slices were resized to 224$\times$224 pixels and z-score normalized for training both the classifier (Module-1) and the detector (Module-2) modules. The segmentator module (Module-3) was pre-trained on ground truth PMBCL ROIs (bounding boxes) and fine-tuned on DLBCL ROIs cropped from the foreground slices and resized to 128$\times$128 pixels. The detailed count for the images in the training, validation and test set for these three tasks are given in Table \ref{tab:data_table}. 
\begin{table}[]
\caption{3D PET axial slices splitting across training, validation and test sets for PMBCL and DLBCL cases for the three modules.} 
\label{tab:data_table}
\begin{center}
\begin{tabular}{@{}llllll@{}}
\toprule
\textbf{Task}                   & \textbf{Dataset}     & \textbf{Train} & \textbf{Valid} & \textbf{Test} & \textbf{Total} \\ \midrule
\multirow{2}{*}{Classification} & PMBCL (axial slices) & 18641          & 6384           & 6101          & 31126          \\
                                & DLBCL (axial slices) & 5786           & 4506           & 2893          & 13185          \\
\hline
\multirow{2}{*}{Detection}      & PMBCL (axial slices) & 1364           & 743            & 851           & 2958           \\
                                & DLBCL (axial slices) & 715            & 261            & 241           & 1217           \\
\hline
\multirow{2}{*}{Segmentation}   & PMBCL (ROIs)         & 1701           & 1050           & 1306          & 4057           \\
                                & DLBCL (ROIs)         & 1034               & 316               & 329              & 1679              \\ 
\bottomrule
\end{tabular}
\end{center}
\end{table}

\subsection{Training the cascaded deep model}
\label{subsec:training}

\begin{enumerate}
\item Module-1 (Slice classifier): To train this module, we used the ImageNet-trained ResNet152 backbone, and pre-trained and tested the fully-connected layers using 31126 axial slices of 3D PET images from the PMBCL cases (for 150 epochs) and subsequently fine-tuned and tested the module using 13185 axial slices from the DLBCL cases (for another 50 epochs).  The minibatch size for training was set to 64 axial images. The model weights were optimized using Adam optimizer with learning rate $= 10^{-5}$, $\beta_1 = 0.9$, $\beta_2 = 0.999$, and $\epsilon = 10^{-8}$. The classification loss function $\mathcal{L}_{classification}$ used is given by, $\mathcal{L}_{classification} = \mathcal{L}_{Focal}$, where $\mathcal{L}_{Focal}$ is the focal loss \cite{8417976} with parameters $\alpha = 0.25$ and $\gamma = 2$. This loss function was chosen to increase the classification accuracy in the presence of class imbalance between the foreground and the background slices.

\item Module-2 (Tumor detector): To train this module, we pre-trained and tested the ImageNet-trained Faster R-CNN network using 2958 foreground axial slices of 3D PET images from the PMBCL cases (for 100 epochs) and then fine-tuned and tested the module using 1217 axial slices from the DLBCL cases (for another 100 epochs). The minibatch size for training was set to 16 axial images. The Adam optimizer was used with the same hyperparameters as used for the classifier module above (except the learning rate was set to $10^{-3}$). The detection loss function $\mathcal{L}_{detection}$ used is given by, $\mathcal{L}_{detection} = \mathcal{L}_{class} + \lambda \mathcal{L}_{box}$, where $\mathcal{L}_{class}$ is the cross-entropy classification loss \cite{crossentropy} for class prediction (tumor), and $\mathcal{L}_{box}$ is the L1 regression loss \cite{crossentropy} for bounding box coordinates prediction. The weight hyperparameter $\lambda = 10.0$ was used during training. 

\item Module-3 (Tumor segmentor): This module was pre-trained and tested on 4057 ground truth PMBCL ROIs (for 25 epochs) and fine-tuned and tested on 1679 DLBCL ROIs (for another 15 epochs) cropped from foreground slices. The minibatch size for training was set to 32 axial images. The segmentation loss function $\mathcal{L}_{segmentation}$ used is given by, $\mathcal{L}_{segmentation} = \mathcal{L}_{generalized Dice} + \lambda \mathcal{L}_{Focal}$, where $\mathcal{L}_{generalized Dice}$ is the generalized Dice loss \cite{Sudre2017} and $\mathcal{L}_{Focal}$ is the focal loss (with $\alpha = 0.25$, $\gamma = 2$). The weight hyperparameter $\lambda = 10.0$ was used during training.
\end{enumerate}

\section{RESULTS}
\label{sec:results}

\begin{figure}
\begin{center}
\includegraphics[scale=0.5]{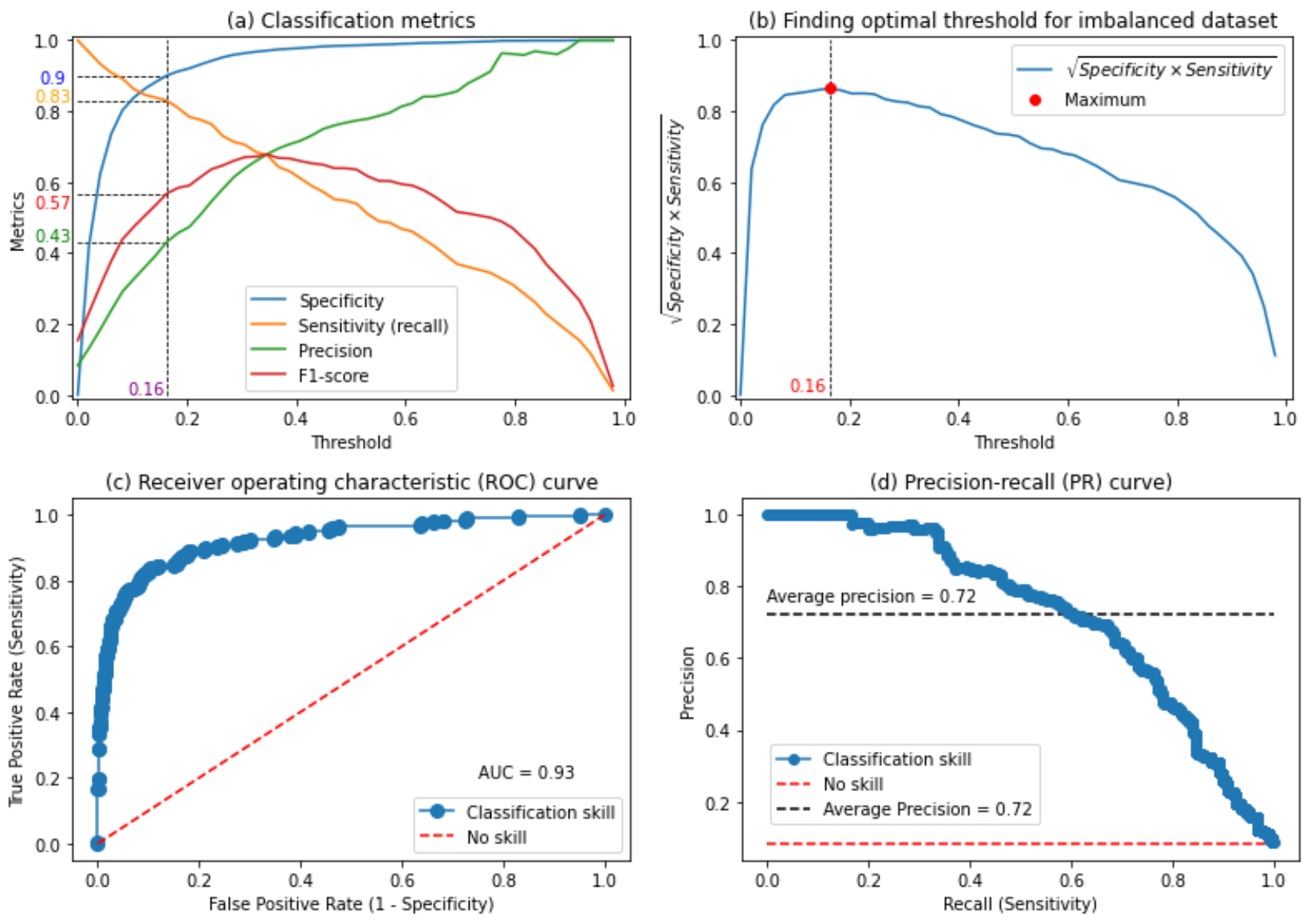}
\end{center}
\caption{Performance of the slice classifier module on the DLBCL test set. (a) Various classification metrics as a function of threshold. (b) The optimal threshold = 0.16 was chosen using the maximum of the geometric mean of specificity and sensitivity. (c) Receiver operating characteristic (ROC) curve, with the area under curve (AUC) = 0.93. (d) Precision-recall (PR) with
average precision = 0.72.} 
\label{fig:classification}
\end{figure} 

\begin{figure}
\begin{center}
\includegraphics[scale=0.6]{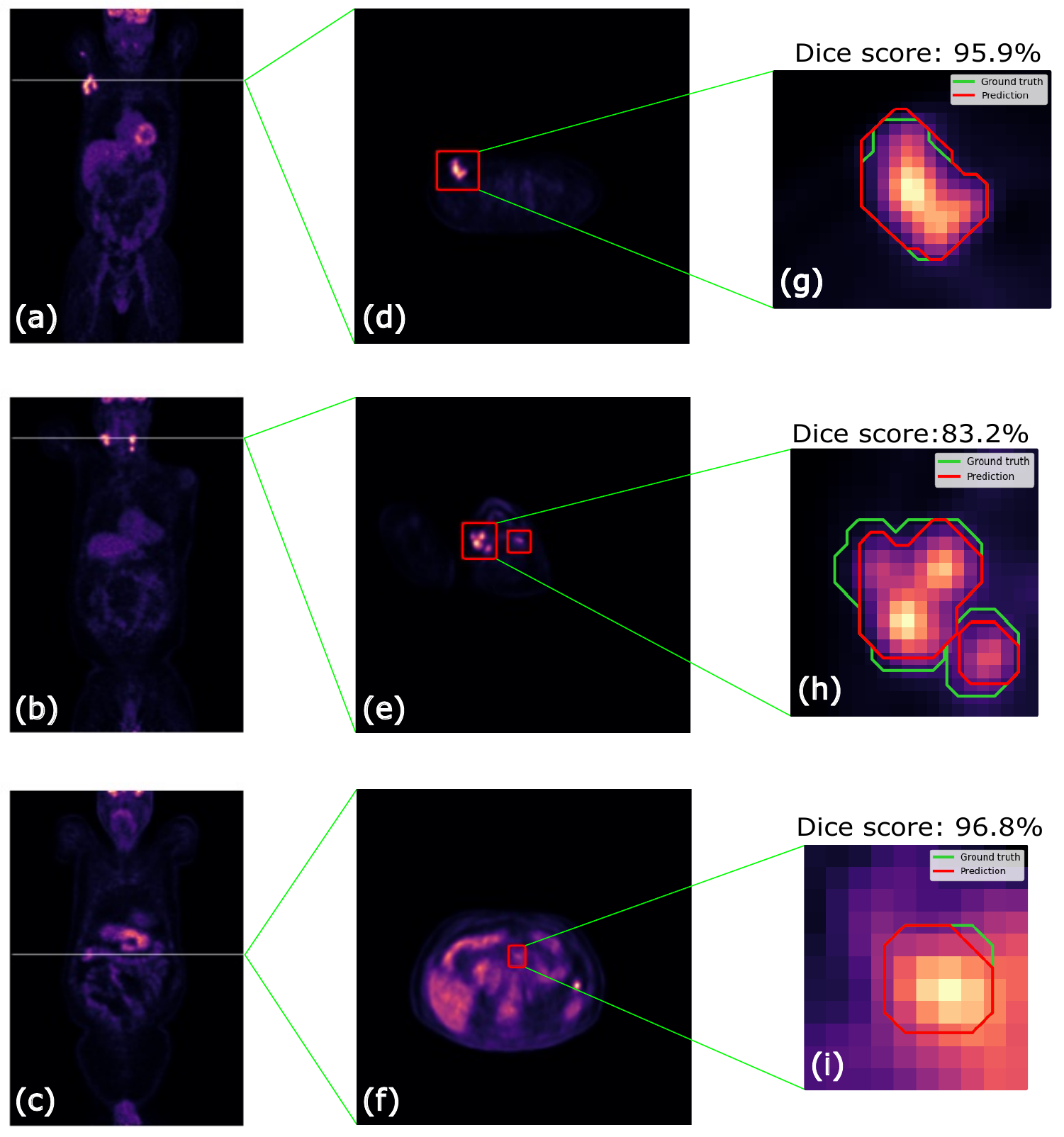}
\end{center}
\caption{Performance of the tumor detector and segmentor modules: (left column) (a)-(c) show 3 representative
DLBCL PET images in the coronal view. (middle column) (d)-(f) show the corresponding selected axial slices (shown as
white horizontal lines in (a)-(c)) with the predicted bounding boxes (ROIs) around the tumors (shown in red). The average detection accuracy on the DLBCL test set was 81\% and average mAP was 0.69. (right column) (g)-(i) show the corresponding ground truth (via physicians, shown in green) and predicted (via our implemented 2D U-Net, shown in red) segmentation contours in 2D, with the average 2D Dice score of 77.9\% $\pm$ 13.2\% across all the slices in the DLBCL test set obtained as output of the detection module. The 2D segmentation contours were aggregated to obtain an
average 3D Dice score of 78.1\% $\pm$ 8.6\% across all test patients.} 
\label{fig:detseg}
\end{figure}

\begin{enumerate}

\item Module 1 (Slice classifier): After pre-training this module on the PMBCL training dataset, a classification accuracy of 89\% and ROC AUC score of 0.92 was obtained on the PMBCL test dataset. Upon fine-tuning the module on DLBCL training set, the module obtained a classification accuracy of 90\% and an ROC AUC score of 0.93. For the DLBCL test dataset containing 2893 axial slices (241 foreground and 2652 background slices), at the optimal threshold = 0.16 (the threshold at which the geometric mean of specificity and sensitivity is maximum, see Figure \ref{fig:classification} (b)), we obtained a specificity of 0.90, sensitivity of 0.83, precision of 0.43, and F1-score of 0.57 (Figure \ref{fig:classification} (a)). The ROC curve (AUC = 0.93) and the precision-recall curve (average precision = 0.72) for the performance of the slice classifier on the DLBCL test dataset are shown in Figure \ref{fig:classification} (c)-(d).

\item Module-2 (Tumor detector): After pre-training and testing this module on only the 2958 foreground slices of the PMBCL test dataset, a tumor detection accuracy of 80\% and mean average precision (mAP) of 0.67 was obtained. Upon subsequent fine-tuning and testing on the 1217 foreground slices of the DLBCL dataset, the trained module obtained a tumor detection accuracy of 81\% and mAP of 0.69 (Table \ref{tab:detection}).

\begin{table}[ht]
\caption{Evaluation of the tumor detector module on the PMBCL and DLBCL test set.} 
\label{tab:detection}
\begin{center}       
\begin{tabular}{@{}lll@{}}
\toprule

\textbf{Test set} & \textbf{Detection accuracy (\%)} & \textbf{Mean average precision (mAP)} \\ \midrule
PMBCL             & 80\%                             & 0.67                                  \\
DLBCL             & \textbf{81\%}                    & \textbf{0.69}                                \\ \bottomrule
\end{tabular}
\end{center}
\end{table}

\item Module-3 (Tumor segmentor): After pre-training and testing on PMBCL ROIs (cropped from foreground slices), the module achieved an average 2D Dice score of 80.7\% $\pm$ 16.0\% on the detected axial slices in the PMBCL test set. Upon aggregating the 2D predictions, a 3D prediction could be generated. An average 3D Dice score of 84.5\% $\pm$ 6.7\% was obtained on the PMBCL test set. After, fine-tuning and testing on DLBCL ROIs, an average 2D Dice score of 77.9\% $\pm$ 13.2\% was achieved on the DLBCL test dataset. Some examples of detection and segmentation tasks on images from the DLBCL test set are shown in Figure \ref{fig:detseg}. An average 3D Dice score of 78.1\% $\pm$ 8.6\% was seen across patients in the DLBCL test dataset. The 3D Dice score obtained using a standard end-to-end (single-step) 3D U-Net architecture on the same DLBCL test set was 58.9\% $\pm$ 16.1\%. This demonstrates that our three-step model performs better than a single-step segmentation network by about 19\% on the same DLBCL test set (Table \ref{tab:segmentation}).

\begin{table}[ht]
\caption{Evaluation of segmentation performance by our proposed model and the 3D U-Net.} 
\label{tab:segmentation}
\begin{center}       
\begin{tabular}{@{}lll@{}}
\toprule
\textbf{Model}                  & \textbf{Test set}      & \textbf{Dice score}  \\ \midrule
\multirow{4}{*}{Proposed model} & \multirow{2}{*}{PMBCL} & 80.7\% $\pm$ 16.0 \% (2D) \\ \cmidrule(l){3-3} 
                                &                        & 84.5\% $\pm$ 6.7\% (3D) \\ \cmidrule(l){2-3} 
                                & \multirow{2}{*}{DLBCL} & 77.9\% $\pm$ 13.2\% (2D) \\ \cmidrule(l){3-3} 
                                &                        & \textbf{78.1\% $\pm$ 8.6\%}(3D)  \\ \midrule
3D U-Net                        & DLBCL                  & 58.9\% $\pm$ 16.1\% (3D) \\ \bottomrule
\end{tabular}
\end{center}
\end{table}

\end{enumerate}

\section{CONCLUSION AND DISCUSSION}
\label{CandD}
Commonly, segmentation of tumors, organs, and other features has been performed using whole-body 3D PET images, either manually, semi-automatically, or via automated deep-learning methods such as the U-Net \cite{pmid32906088}. Automated deep learning methods working on whole-body PET images have so far not been accurate enough \cite{pmid24845019, RIZWANIHAQUE2020100297}.

In this work, we have presented a fully automated three-step deep model for fast and efficient segmentation of tumors from the DLBCL PET images. Our supervised slice classifier and tumor detector modules help improve the overall network
segmentation performance since they localize the tumors for the segmentor module to delineate. The slice classifier module also helps eliminate the need to train the detector module on tumor-less background slices. Moreover, as compared to direct deep learning-based segmentation of tumors in whole-body PET images (U-Net 3D Dice score = 58.9\% $\pm$ 16.1\%), our three-step model shows improved performance (3D Dice score = 78.1\% $\pm$ 8.6\%) by about 19\%.  Another reason for the performance boost can also be attributed to the fact that in our method each specialized module performing a specific task can be trained independently to a high degree of skill, rather than a single network with suboptimal performance on overall segmentation (unless a very large dataset is provided). 

As the performance of the overall network largely depended on the performance of the slice classifier and the tumor detector modules (perfect tumor detection leads to highly accurate segmentation prediction), in the future we would like to explore various methods for improving the performance of our detection module, such as by incorporating deformable convolutions units into the detection network \cite{dcn}. Overall, deep learning-based segmentation methods for oncological PET images hold significant potential towards personalized cancer therapy in the future.

\acknowledgments 
 
This work is supported by the Canadian Institutes of Health Research (CIHR) Project Grant PJT-173231. S.A. would also like to acknowledge the support from SNMMI Medical \& Science Student Research grant 2021 and the Education and Research Foundation for Nuclear Medicine and Molecular Imaging.

\bibliography{report} 

\begin{thebibliography}{10}

\bibitem{pmid9166827}
``{{A} clinical evaluation of the {I}nternational {L}ymphoma {S}tudy {G}roup classification of non-{H}odgkin's lymphoma. {T}he {N}on-{H}odgkin's {L}ymphoma {C}lassification {P}roject},'' {\em Blood}~{\bf 89},  3909--3918 (Jun 1997).

\bibitem{pmid33315178}
Weisman, A.~J., Kim, J., Lee, I., McCarten, K.~M., Kessel, S., Schwartz, C.~L., Kelly, K.~M., Jeraj, R., Cho, S.~Y., and Bradshaw, T.~J., ``{{A}utomated quantification of baseline imaging {P}{E}{T} metrics on {F}{D}{G} {P}{E}{T}/{C}{T} images of pediatric {H}odgkin lymphoma patients},'' {\em EJNMMI Phys}~{\bf 7},  76 (Dec 2020).

\bibitem{pmid28971313}
Slattery, A., ``{{V}alidating an image segmentation program devised for staging lymphoma},'' {\em Australas Phys Eng Sci Med}~{\bf 40},  799--809 (Dec 2017).

\bibitem{pmid32532925}
Capobianco, N., Meignan, M., Cottereau, A.~S., Vercellino, L., Sibille, L., Spottiswoode, B., Zuehlsdorff, S., Casasnovas, O., Thieblemont, C., and Buvat, I., ``{{F}-{F}{D}{G} {U}ptake {C}lassification {E}nables {T}otal {M}etabolic {T}umor {V}olume {E}stimation in {D}iffuse {L}arge {B}-{C}ell {L}ymphoma},'' {\em J Nucl Med}~{\bf 62},  30--36 (01 2021).

\bibitem{pmid30625203}
Guo, B., Tan, X., Ke, Q., and Cen, H., ``{{P}rognostic value of baseline metabolic tumor volume and total lesion glycolysis in patients with lymphoma: {A} meta-analysis},'' {\em PLoS One}~{\bf 14}(1),  e0210224 (2019).

\bibitem{pmid31978225}
Vercellino, L., Cottereau, A.~S., Casasnovas, O., Tilly, H., Feugier, P., Chartier, L., Fruchart, C., Roulin, L., Oberic, L., Pica, G.~M., Ribrag, V., Abraham, J., Simon, M., Gonzalez, H., Bouabdallah, R., Fitoussi, O., Sebban, C., López-Guillermo, A., Sanhes, L., Morschhauser, F., Trotman, J., Corront, B., Choufi, B., Snauwaert, S., Godmer, P., Briere, J., Salles, G., Gaulard, P., Meignan, M., and Thieblemont, C., ``{{H}igh total metabolic tumor volume at baseline predicts survival independent of response to therapy},'' {\em Blood}~{\bf 135},  1396--1405 (04 2020).

\bibitem{pmid32724136}
Martín-Saladich, Q., Reynés-Llompart, G., Sabaté-Llobera, A., Palomar-Muñoz, A., Domingo-Domènech, E., and Cortés-Romera, M., ``{{C}omparison of different automatic methods for the delineation of the total metabolic tumor volume in {I}-{I}{I} stage {H}odgkin {L}ymphoma},'' {\em Sci Rep}~{\bf 10},  12590 (07 2020).

\bibitem{tmtv_1}
Casasnovas, R.-O., Kanoun, S., Tal, I., Cottereau, A.-S., Edeline, V., Brice, P., Bouabdallah, R., Salles, G.~A., Stamatoullas, A., Dupuis, J., Reman, O., Gastinne, T., Joly, B., Bouabdallah, K., Nicolas-Virelizier, E., Andre, M., Mounier, N., Ferme, C., Meignan, M., and Berriolo-Riedinger, A., ``Baseline total metabolic volume (tmtv) to predict the outcome of patients with advanced hodgkin lymphoma (hl) enrolled in the ahl2011 lysa trial.,'' {\em Journal of Clinical Oncology}~{\bf 34}(15\_suppl),  7509--7509 (2016).

\bibitem{pattern_recog}
Meyer-Base, A., Schmid, V., and Books, E. A.~A.,  [{\em Pattern recognition and signal analysis in medical imaging}{\nolinebreak\hspace{0.1em}]}, Elsevier/Academic Press, Waltham, MA;Oxford, UK;, second~ed. (2014).

\bibitem{pmid32068507}
Willemink, M.~J., Koszek, W.~A., Hardell, C., Wu, J., Fleischmann, D., Harvey, H., Folio, L.~R., Summers, R.~M., Rubin, D.~L., and Lungren, M.~P., ``{{P}reparing {M}edical {I}maging {D}ata for {M}achine {L}earning},'' {\em Radiology}~{\bf 295},  4--15 (04 2020).

\bibitem{LUNDERVOLD2019102}
Lundervold, A.~S. and Lundervold, A., ``An overview of deep learning in medical imaging focusing on mri,'' {\em Zeitschrift fur Medizinische Physik}~{\bf 29}(2),  102--127 (2019).
\newblock Special Issue: Deep Learning in Medical Physics.

\bibitem{pmid33771905}
Buvat, I. and Orlhac, F., ``{{T}he {T}.{R}.{U}.{E}. {C}hecklist for {I}dentifying {I}mpactful {A}rtificial {I}ntelligence-{B}ased {F}indings in {N}uclear {M}edicine: {I}s {I}t {T}rue? {I}s {I}t {R}eproducible? {I}s {I}t {U}seful? {I}s {I}t {E}xplainable?},'' {\em J Nucl Med}~{\bf 62},  752--754 (06 2021).

\bibitem{pmid34537131}
Yousefirizi, F., Jha, A.~K., Brosch-Lenz, J., Saboury, B., and Rahmim, A., ``{{T}oward {H}igh-{T}hroughput {A}rtificial {I}ntelligence-{B}ased {S}egmentation in {O}ncological {P}{E}{T} {I}maging},'' {\em PET Clin}~{\bf 16},  577--596 (Oct 2021).

\bibitem{unet}
Ronneberger, O., Fischer, P., and Brox, T.,  [{\em U-Net: Convolutional Networks for Biomedical Image Segmentation}{\nolinebreak\hspace{0.1em}]},  234--241, Springer International Publishing, Cham (2015).

\bibitem{maskrcnn}
He, K., Gkioxari, G., Dollar, P., and Girshick, R., ``Mask r-cnn,'' {\em IEEE transactions on pattern analysis and machine intelligence}~{\bf 42}(2),  386--397 (2020).

\bibitem{pmid31217099}
Kumar, A., Fulham, M., Feng, D., and Kim, J., ``{{C}o-{L}earning {F}eature {F}usion {M}aps from {P}{E}{T}-{C}{T} {I}mages of {L}ung {C}ancer},'' {\em IEEE Trans Med Imaging}  (Jun 2019).

\bibitem{pmid29652908}
Blanc-Durand, P., Van Der~Gucht, A., Schaefer, N., Itti, E., and Prior, J.~O., ``{{A}utomatic lesion detection and segmentation of 18{F}-{F}{E}{T} {P}{E}{T} in gliomas: {A} full 3{D} {U}-{N}et convolutional neural network study},'' {\em PLoS One}~{\bf 13}(4),  e0195798 (2018).

\bibitem{det_seg_1}
Zhu, Z., Jin, D., Yan, K., Ho, T.-Y., Ye, X., Guo, D., Chao, C.-H., Xiao, J., Yuille, A., and Lu, L.,  [{\em Lymph Node Gross Tumor Volume Detection and Segmentation via Distance-Based Gating Using 3D CT/PET Imaging in Radiotherapy}{\nolinebreak\hspace{0.1em}]},  753--762, Springer International Publishing, Cham (2020).

\bibitem{5459262}
Lempitsky, V., Kohli, P., Rother, C., and Sharp, T., ``Image segmentation with a bounding box prior,'' in [{\em 2009 IEEE 12th International Conference on Computer Vision}{\nolinebreak\hspace{0.1em}]},   277--284 (2009).

\bibitem{pmid33937842}
Weisman, A.~J., Kieler, M.~W., Perlman, S.~B., Hutchings, M., Jeraj, R., Kostakoglu, L., and Bradshaw, T.~J., ``{{C}onvolutional {N}eural {N}etworks for {A}utomated {P}{E}{T}/{C}{T} {D}etection of {D}iseased {L}ymph {N}ode {B}urden in {P}atients with {L}ymphoma},'' {\em Radiol Artif Intell}~{\bf 2},  e200016 (Sep 2020).

\bibitem{7780459}
He, K., Zhang, X., Ren, S., and Sun, J., ``Deep residual learning for image recognition,'' in [{\em 2016 IEEE Conference on Computer Vision and Pattern Recognition (CVPR)}{\nolinebreak\hspace{0.1em}]},   770--778 (2016).

\bibitem{5206848}
Deng, J., Dong, W., Socher, R., Li, L.-J., Li, K., and Fei-Fei, L., ``Imagenet: A large-scale hierarchical image database,'' in [{\em 2009 IEEE Conference on Computer Vision and Pattern Recognition}{\nolinebreak\hspace{0.1em}]},   248--255 (2009).

\bibitem{pmid27295650}
Ren, S., He, K., Girshick, R., and Sun, J., ``{{F}aster {R}-{C}{N}{N}: {T}owards {R}eal-{T}ime {O}bject {D}etection with {R}egion {P}roposal {N}etworks},'' {\em IEEE Trans Pattern Anal Mach Intell}~{\bf 39},  1137--1149 (06 2017).

\bibitem{fpn}
Lin, T.-Y., Dollár, P., Girshick, R., He, K., Hariharan, B., and Belongie, S., ``Feature pyramid networks for object detection,'' (2016).

\bibitem{consensus}
Yousefirizi, F., Bloise, I., Martineau, P., Wilson, D., Benard, F., Bradshaw, T., Rahmim, A., and Uribe, C.~F., ``Reproducibility of a semi-automatic gradient-based segmentation approach for lymphoma pet,'' {\em European Journal of Nuclear Medicine and Molecular Imaging}~{\bf 48}(1) (2021).

\bibitem{8417976}
Lin, T.-Y., Goyal, P., Girshick, R., He, K., and Dollár, P., ``Focal loss for dense object detection,'' {\em IEEE Transactions on Pattern Analysis and Machine Intelligence}~{\bf 42}(2),  318--327 (2020).

\bibitem{crossentropy}
Murphy, K.~P. and Central, E.,  [{\em Machine learning: a probabilistic perspective}{\nolinebreak\hspace{0.1em}]}, MIT Press, Cambridge, MA (2012).

\bibitem{Sudre2017}
Sudre, C.~H., Li, W., Vercauteren, T., Ourselin, S., and Jorge~Cardoso, M., ``Generalised dice overlap as a deep learning loss function for highly unbalanced segmentations,'' {\em Deep learning in medical image analysis and multimodal learning for clinical decision support : Third International Workshop, DLMIA 2017, and 7th International Workshop, ML-CDS 2017, held in conjunction with MICCAI 2017 Quebec City, QC,...}~{\bf 2017},  240--248 (2017).
\newblock 34104926[pmid].

\bibitem{pmid32906088}
Weisman, A.~J., Kieler, M.~W., Perlman, S., Hutchings, M., Jeraj, R., Kostakoglu, L., and Bradshaw, T.~J., ``{{C}omparison of 11 automated {P}{E}{T} segmentation methods in lymphoma},'' {\em Phys Med Biol}~{\bf 65},  235019 (11 2020).

\bibitem{pmid24845019}
Foster, B., Bagci, U., Mansoor, A., Xu, Z., and Mollura, D.~J., ``{{A} review on segmentation of positron emission tomography images},'' {\em Comput Biol Med}~{\bf 50},  76--96 (Jul 2014).

\bibitem{RIZWANIHAQUE2020100297}
{Rizwan I Haque}, I. and Neubert, J., ``Deep learning approaches to biomedical image segmentation,'' {\em Informatics in Medicine Unlocked}~{\bf 18},  100297 (2020).

\bibitem{dcn}
Dai, J., Qi, H., Xiong, Y., Li, Y., Zhang, G., Hu, H., and Wei, Y., ``Deformable convolutional networks,''  764--773, IEEE (2017).

\end{thebibliography}
\bibliographystyle{spiebib} 

\end{document}